%% file: ICRC2023_template_IceCube.tex
\documentclass[a4paper,11pt]{article}

\usepackage{pos}
\usepackage{hyperref}
\usepackage{siunitx}

\def\apj{\ref@jnl{ApJ}}

\title{Searching for high-energy neutrinos from shock-interaction powered supernovae with the IceCube Neutrino Observatory}

\ShortTitle{IceCube search of neutrinos from interaction-powered supernovae}

\author{The IceCube Collaboration \\{\normalsize \normalfont(a complete list of authors can be found at the end of the proceedings)}\\}

\emailAdd{lincetto@astro.ruhr-uni-bochum.de}
\abstract{

The sources of the astrophysical neutrino flux discovered by IceCube are for the most part unresolved. Extragalactic core-collapse supernovae (CCSNe) have been suggested as candidate multi-messenger sources. In interaction-powered supernovae, a shock propagates in a dense circumstellar medium (CSM), producing a bright optical emission and potentially accelerating particles to relativistic energies. Shock interaction is believed to be the main energy source for Type IIn supernovae (identified by narrow lines in the spectrum), hydrogen-rich superluminous supernovae and a subset of hydrogen-poor superluminous supernovae. Production of high-energy neutrinos is expected in collisions between the accelerated protons in the shocks and the cold CSM particles. We select a catalog of interaction-powered supernovae from the Bright Transient Survey of the Zwicky Transient Facility. We exploit a novel modeling effort that connects the time evolution of the optical emission to the properties of the ejecta and the CSM, allowing us to set predictions of the neutrino flux for each source. In this contribution, we describe a stacking search for high-energy neutrinos from this population of CCSNe with the IceCube Neutrino Observatory.

\vspace{4mm}
{\bfseries Corresponding authors:}
Massimiliano Lincetto$^{1*}$\\
{$^{1}$ \itshape Fakultät für Physik \& Astronomie, Ruhr-Universität Bochum, D-44780 Bochum, Germany}\\
$^*$ Presenter

\ConferenceLogo{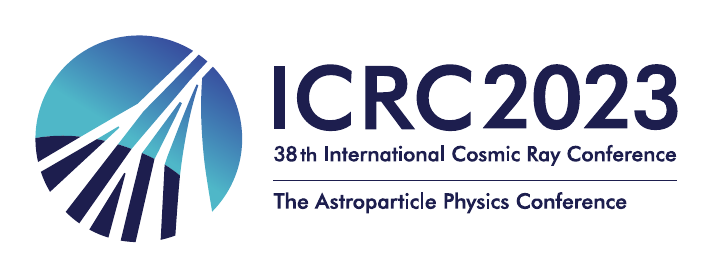}

\FullConference{The 38th International Cosmic Ray Conference (ICRC2023)\\ 26 July -- 3 August, 2023\\ Nagoya, Japan}
}

\begin{document}

\maketitle

\section{Introduction}
\label{sec1}

The IceCube Neutrino Observatory is a large-volume neutrino detector located at the Geographic South Pole. IceCube consists of an array of photomultipliers submerged in the Antarctic ice, detecting the Cherenkov light induced by charged secondary particles of high-energy neutrino interactions \cite{IceCube:2016zyt}.
Astrophysical neutrinos point back to their sources and are unequivocal signatures of hadronic acceleration. They originate from the decay of charged mesons produced in interactions between accelerated particles and matter or radiation in cosmic-ray sources. Neutral mesons produced in the same processes decay to pairs of high-energy gamma rays. IceCube has discovered and characterised an astrophysical flux of high-energy neutrinos~\cite{IceCube:2016umi, IceCube:2020acn, IceCube:2020wum}. Its origin  is for the large part unknown. High-energy neutrino emission has been associated with individual transient and steady sources, such as the flaring blazar TXS0506+056~\cite{IceCube:2018dn-TXS}
and the nearby active galaxy NGC 1068~\cite{IceCube:2022der}, as well as the Galactic Plane~\cite{IceCube-GP}. However, no significant excess from a population of sources has been observed, and the majority of the flux observed by IceCube is still unexplained. The energy density of astrophysical neutrinos is comparable to the one observed in high-energy gamma rays and ultra-high-energy cosmic rays. However, the multimessenger evidence suggests that if all the sources of neutrinos were to emit proportionally in gamma rays, the resulting extragalactic gamma background would overshoot the diffuse extragalactic gamma flux limits set by Fermi-LAT \cite{Murase:2015xka}. This means that a significant fraction of astrophysical neutrinos must originate in sources that are gamma-ray faint. Different classes of optically-observed astronomical transients have been suggested as potential neutrino emitters. Among these are tidal disruption events \cite{Stein:2020xhk}, choked-jet supernovae (core-collapse supernovae of type Ib and Ic) and interacting supernovae (core-collapse supernovae of type IIn and superluminous supernovae) \cite{Murase:2017pfe}. A previous search for neutrinos from core-collapse supernovae with IceCube has not found any excess signal from such a population~\cite{IceCube:2023esf}. Some supernovae belong to the category of shock-powered transients, primarily observed in the optical bands \cite{Fang:2020bkm}. They are characterized by strong interaction of their ejecta with a dense circumstellar medium. This produces narrow lines in their spectrum (as in the case of supernovae of type IIn) or exceptional brightness (as in the case of hydrogen-rich superluminous supernovae, or SLSN-II). In this work, we present a study specifically targeted to supernovae which are likely interaction powered. A search for high-energy neutrinos from such sample could help in constraining the shock-interaction models for hadronic particle acceleration.

\section{Neutrinos from interaction-powered supernovae}
\label{sec:sample}%
\begin{figure}
    \centering
    \includegraphics[width=0.55\textwidth]{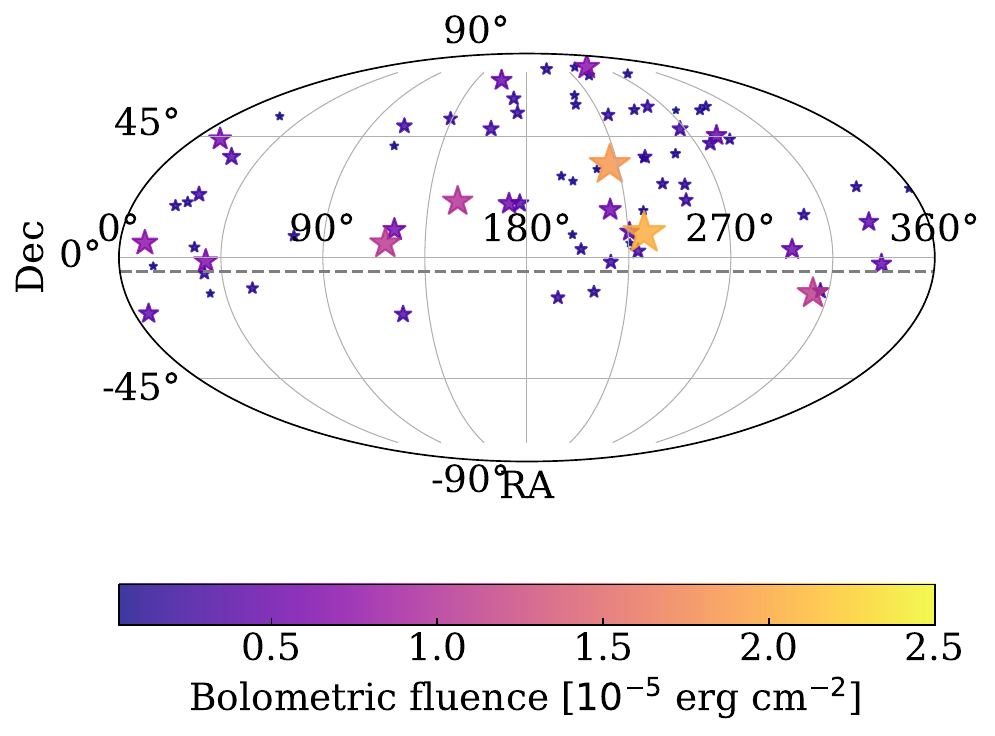}
    \caption{Representation of the interaction-powered supernovae catalog in equatorial coordinates, consisting of 74 sources, among which 65 in the Northern hemisphere. Marker sizes are proportional to their estimated bolometric flux at Earth, calculated as described in Sec.~\ref{sec:sample}. The horizontal dashed line represents the separation between the two hemispheres that applies to IceCube neutrino data, see Sec.~\ref{sec:neutrino-data}.}
    \label{fig:catalogue}
\end{figure}

Core-collapse supernovae (CCSNe), ubiquitous in our Universe, convert large amounts of gravitational energy into low-energy neutrinos, thermal photons and kinetic energy of the ejecta. An especially luminous subset of hydrogen-rich CCSNe is represented by Type IIn SNe, characterised by the presence of narrow lines in their optical spectrum. This feature is a signature of intense interaction between the supernova ejecta and a dense circumstellar medium (CSM). Radiation generated in the interaction process can explain their observed high luminosities in optical bands. This is especially the case for hydrogen-rich superluminous supernovae (SLSN-II), considered to be the most luminous examples of interaction-powered supernovae and hence an extension of the IIn population~\cite{Gal-Yam:2018out}. In interaction-powered supernovae, shocks produced as the ejecta crash into the CSM may be able to accelerate particles to relativistic energies. The accelerated protons colliding with protons and nuclei of the cold CSM could produce high-energy neutrinos~\cite{Pitik:2021dyf}. State-of-the-art models can account for the observed photometric evolution of the supernova in order to infer the properties of the ejecta and CSM configuration. In particular, sources with high peak luminosities and intermediate rise times are the most favoured for production of high-energy neutrinos~\cite{Pitik:2023vcg}. To build a sample of interaction-powered supernovae, we select type IIn supernovae and type II superluminous supernovae from the Bright Transient Survey of the Zwicky Transient Facility \cite{Fremling:2019dvl, Perley:2020ajb}. We first select events with robust spectroscopic classification, for which the time evolution has been well-observed in both bands, allowing for a reliable estimation of the rise and peak times of the supernova. The final sample consists of 74 sources detected up to May 31, 2021. The sources' distribution in the sky is shown in Fig.~\ref{fig:catalogue}. 65 sources are located in the Northern hemisphere.
We retrieve forced-photometry light curves measured in the ZTF red (ZTF-r) and green (ZTF-g) passbands and we correct the measured fluxes for Galactic extinction \cite{SFDMap}. Since the measurements in the two bands are not simultaneous, the individual light curves are linearly interpolated to estimate the value of the ZTF-r flux in correspondence of each ZTF-g measurement, and vice-versa. At each time of measurement, the flux is integrated between the center wavelengths of the two filters using a simple trapezoid method, in order to estimate the source instantanous luminosity. This estimation, defined in the following as "pseudo-bolometric luminosity", is taken as a lower limit to the bolometric emission of the supernova. The obtained luminosity as a function of time is finally interpolated using Gaussian process regression~\cite{george}. The duration of the optical emission, to be used as a time window for the neutrino search described in Sec.~\ref{sec:stacking}, is estimated from the photometric data. The start of the time window is defined as the time of first detection in either band or, if a non detection is reported in the 20 days before, as the average between the last non-detection and the first detection. The end of the time window is defined as the end of the phase in which the supernova is detected in both bands, corresponding to the last time at which the "pseudo-bolometric luminosity" can be estimated reliably. Time windows extending beyond May 31, 2021 (end of the IceCube 2020-21 data season) are artificially cut at the said date.

\section{Neutrino data samples}
\label{sec:neutrino-data}
IceCube searches for point-like sources rely on samples of candidate muon neutrino and antineutrino events. We consider two track-like event selections with comparable detector livetime up to the end of the 2020-21 data season. The first ("point source tracks", from hereon "PS") covers the full sky and has been widely used in previous point-source searches \cite{IceCube:2019cia}. The background of this sample consists of mainly atmospheric neutrinos in the Northern hemisphere (upgoing events) and atmospheric muons in the Southern hemisphere (downgoing events). This requires different selection criteria for the two hemispheres, that are conventionally separated at a declination of \SI{-5}{deg}. For this selection, the background-like datasets to be used in the analysis are generated by shuffling the events in time and randomizing their right ascension coordinate. The second sample is a refined selection of tracks restricted to the Northern hemisphere ("northern tracks", from hereon "NT") with improved estimators for energy and directional reconstruction as described in~\cite{IceCube:2022der}. This sample shows an excellent agreement between data and simulation, hence a pure Monte Carlo data set can be used to generate background-like data samples, under the assumption of constant rate. Since the supernovae discovered by ZTF are mostly in the Northern hemisphere, it can be expected for the northern data sample to provide a better sensitivity compared to the all sky selection.

\section{Stacking analysis}
\label{sec:stacking}
\begin{figure}
    \centering
    \includegraphics[width=0.45\textwidth]{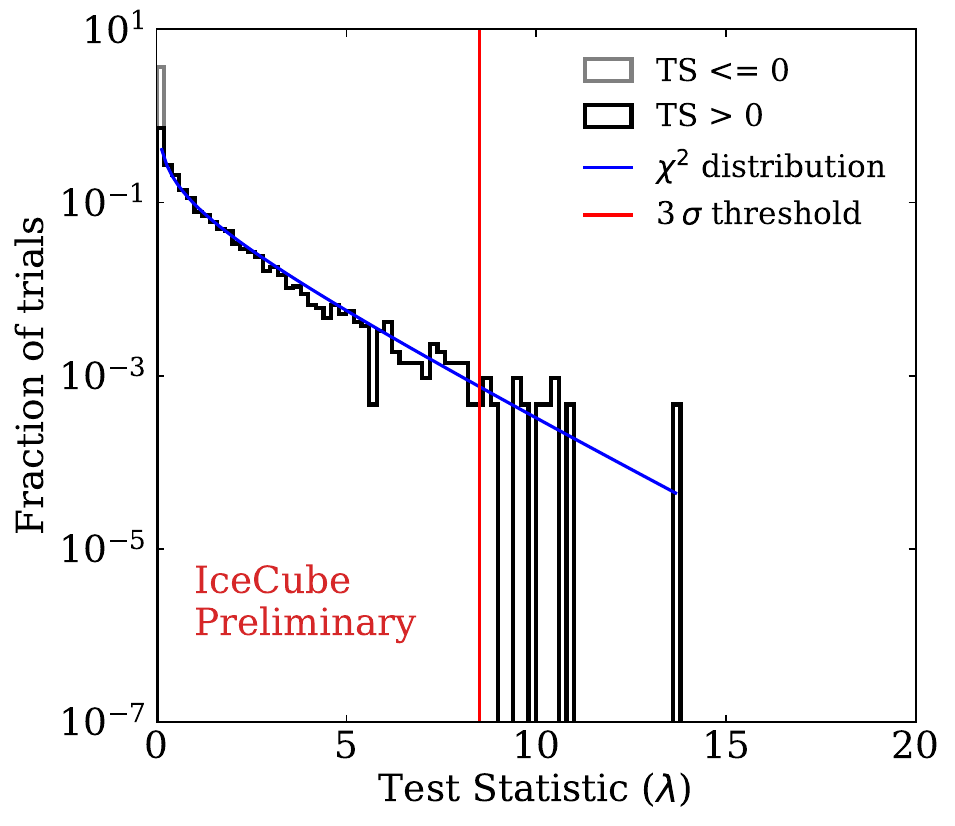}%
    \includegraphics[width=0.45\textwidth]{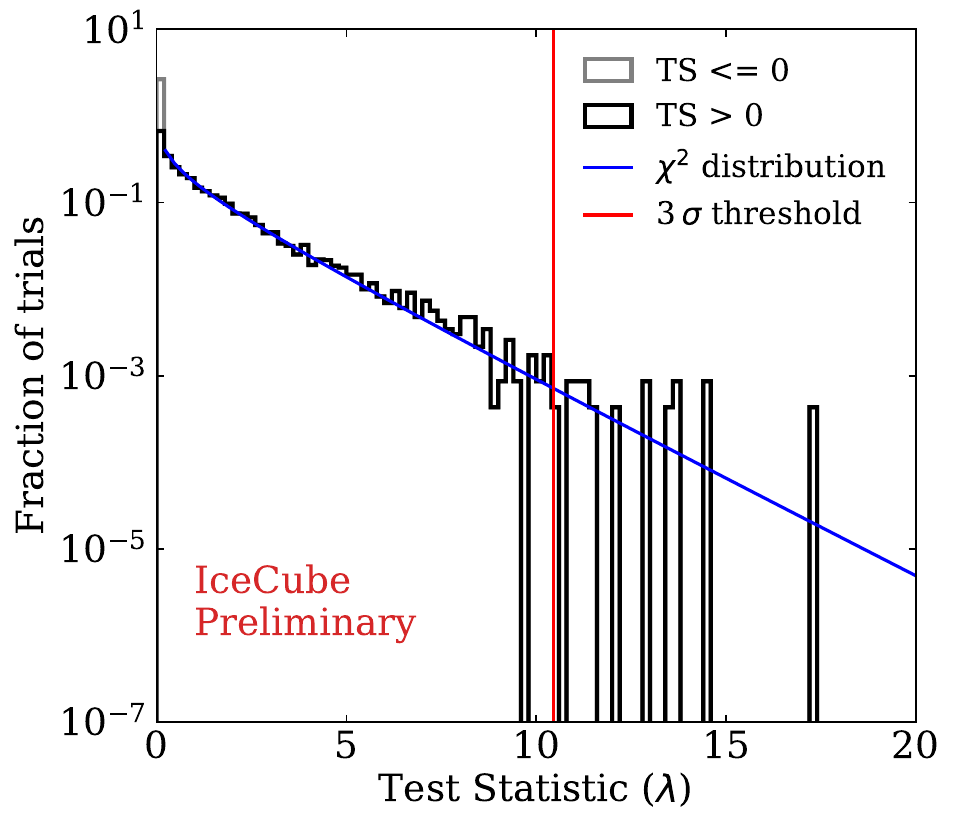}%
    \caption{Background test statistic distributions from pseudoexperiments. Left: complete source catalog with the all sky track sample described in Ref.~\cite{IceCube:2019cia} where the background-like pseudoexperiments are obtained by shuffling experimental data. Right: Northern sources catalog with the Northern sky track sample described in Ref.~\cite{IceCube:2022der}, where the background-like pseudoexperiments are obtaiened from Monte Carlo simulations.}
    \label{fig:TSdist}
\end{figure}

\begin{figure}
    \centering
    \includegraphics[width=0.45\textwidth]{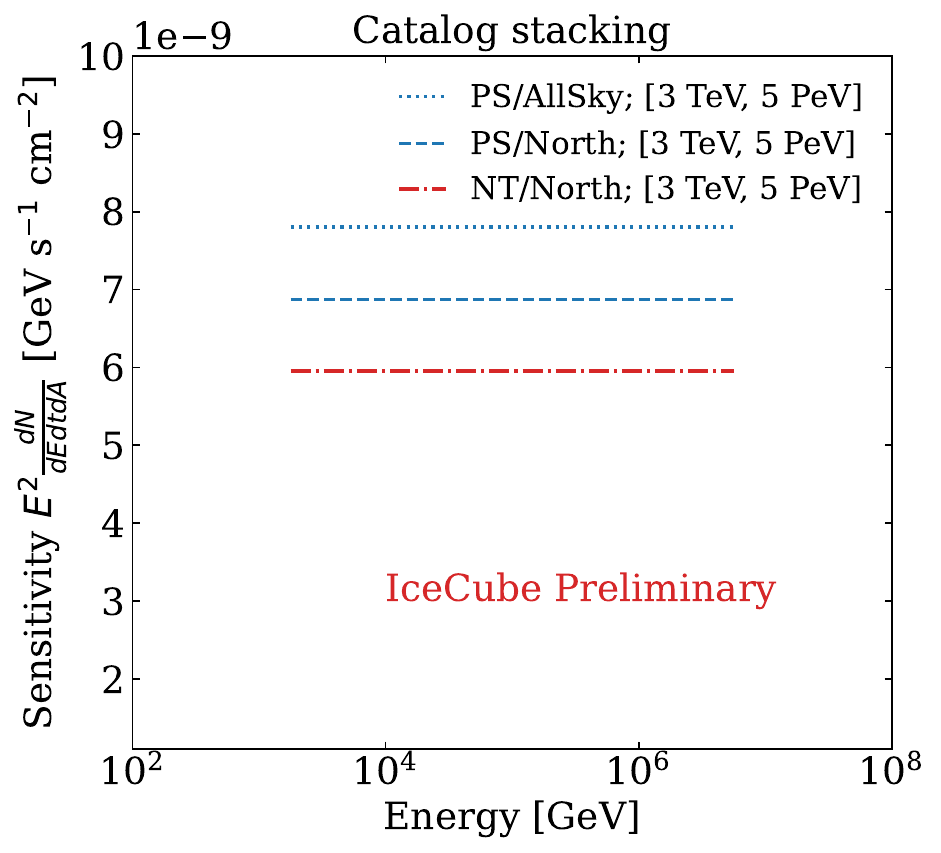}%
    \includegraphics[width=0.45\textwidth]{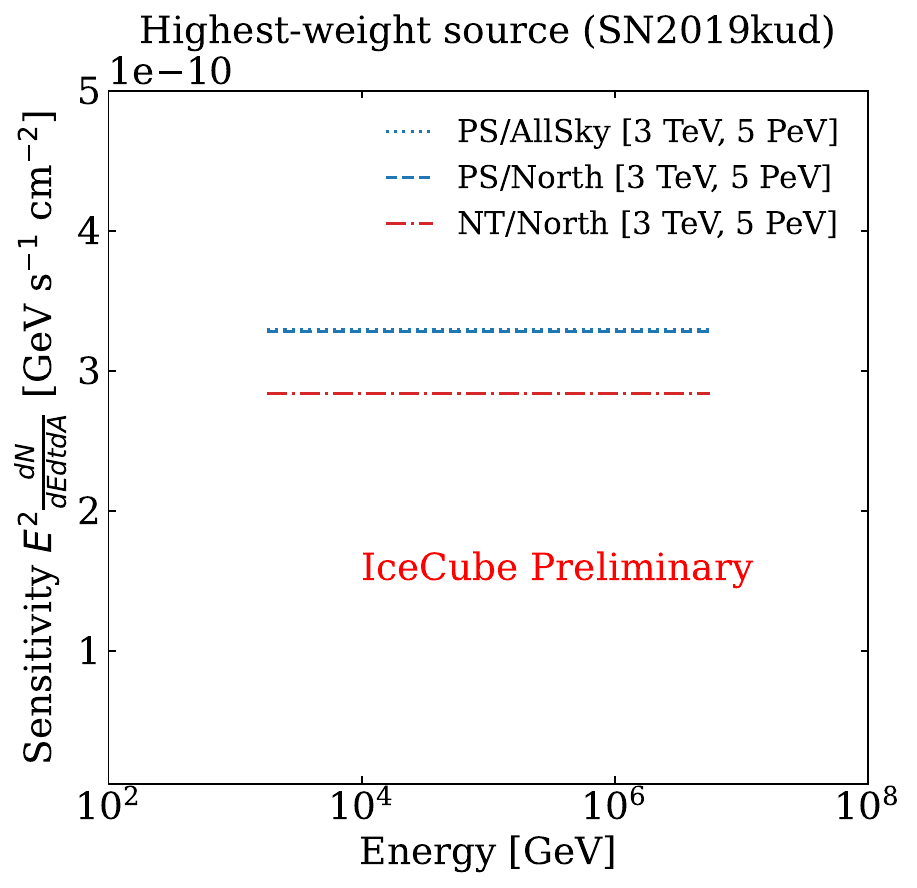}%
    \caption{Sensitivity (flux normalization at $\SI{1}{GeV}$) estimated with the two considered data samples for a spectral index $\gamma = 2.0$. "PS" indicates the all sky track sample described in Ref.~\cite{IceCube:2019cia}, "NT" the Northern sky track sample described in Ref.~\cite{IceCube:2022der} (see Sec.~\ref{sec:neutrino-data}). Both samples are evaluated against the subset of sources in the Northern hemisphere ("North"). The PS sample is also evaluated against the complete catalog ("AllSky"). The sensitivities are shown for the sum of all sources (left) and for the source with the highest weight in the sample (right). The sensitivities are defined for the energy range from $\sim \SI{3}{TeV}$ to $\sim \SI{5}{PeV}$, corresponding to the 90\% sensitive range estimated from the all sky analysis.}
    \label{fig:stacking}
\end{figure}

\begin{figure}
    \centering
    \includegraphics[width=0.45\textwidth]{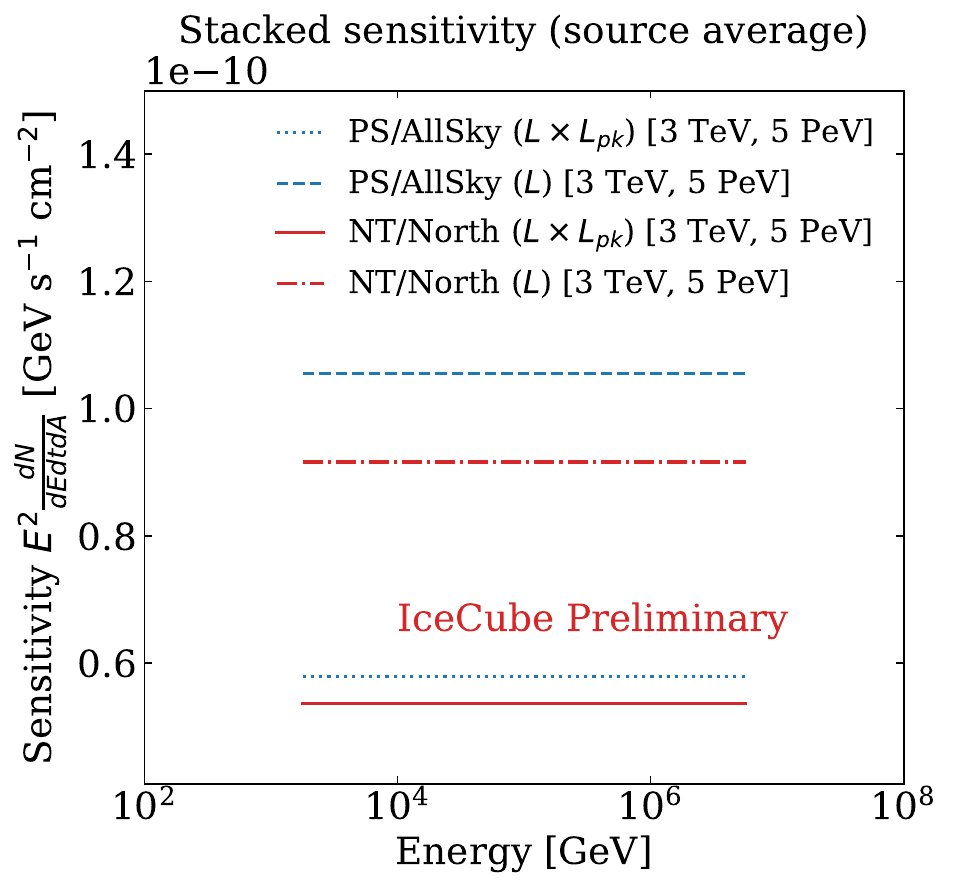}%
    \caption{Sensitivity (flux normalization at $\SI{1}{GeV}$) for a spectral index $\gamma = 2.0$ estimated from the all sky track sample described in Ref.~\cite{IceCube:2019cia} ("PS") for the complete catalog ("AllSky") and from the Northern sky track sample described in Ref.~\cite{IceCube:2022der} ("NT") for the subset of sources in the Northern hemisphere ("North"). The sensitivity is divided by the number of sources in each catalog and represented for the two weighting schemes: average luminosity ($L$) and average luminosity multiplied by the peak luminosity ($L \times L_{pk}$). The sensitivities are defined for the energy range from $\sim \SI{3}{TeV}$ to $\sim \SI{5}{PeV}$, corresponding to the 90\% sensitive range estimated from the all sky analysis.}
    \label{fig:stacking-weights}
\end{figure}

The chosen analysis method for the search of an excess of neutrinos from the supernova catalog is an unbinned likelihood stacking. The likelihood function is defined as follows:
\begin{equation}
    \mathcal{L}(n_s, \gamma) = \prod _{i=0} \left[ \frac{n_s}{N} \sum_{j=0} ^M w_j \mathcal{S}_j(\theta_{i}, \gamma)  + \left( 1- \frac{n_{s}}{N} \right) \mathcal{B}(\theta_i)  \right]
\end{equation}
where $\theta_i$ are the properties (direction, direction error estimator, energy) of the $i$-th neutrino in the sample (out of $N$), $n_s$ is the parameter representing the total number of signal events from the source catalog, $w_j$ is the injection weight assigned to source $j$ (out of $M$). $\mathcal{S}$ and $\mathcal{B}$ are the signal and background probability density functions (PDFs), $\gamma$ is represents the spectral index of the signal flux. The signal and background PDFs consists of a spatial component, a temporal component and an energy component. The background PDF is constant over time and right ascension, depending only on the source declination and energy. The signal temporal PDF for a source $j$ is taken as uniform over the emission time of the source. Similarly to previous analyses \cite{IceCube:2023esf}, the signal energy PDF is an unbroken power law with spectral index $\gamma$. The spatial component of the signal PDF is a circular, two-dimensional, Gaussian ditribution evaluated on the angular separation between the neutrino direction and the source position. The width of the Gaussian is, for each individual event, given by the angular error estimator associated with the reconstructed direction of the neutrino. Concerning the "NT" data set, we note that while this work makes use of the improved estimators of event observables, it does not adopt the improved PDF modeling used in the NGC 1068 analysis \cite{IceCube:2022der}. The test statistic (TS) is defined as the likelihood ratio of the signal plus background $\mathcal{L}(n_s, \gamma)$ to the background-only $\mathcal{L}(n_s = 0)$ hypotheses. The background TS distribution is evaluated by simulating a large number of background-like data samples (pseudoexperiments), and evaluating the TS corresponding to the maximum of the likelihood ratio in the $(n_s,\gamma)$ parameter space. The background TS distributions for the two considered data samples are shown in Fig.\ref{fig:TSdist}.
To estimate the analysis sensitivity, the evaluation is repeated for pseudoexperiments where a simulated signal is injected on top of the background. The signal neutrinos are distributed on the source catalog according to the injection weights, $w_j$:
\begin{equation}
    w_j \propto \frac{w^{*}_j}{4 \pi d^2_L} T_j \int_{E_{\mathrm{min}}}^{E_{\mathrm{max}}} \frac{dN}{dE}(E) A_{\mathrm{eff}}(\delta_j, E)\,dE
\end{equation}
where $w^{*}_j$ is the intrinsic source weight, $d_L$ is the luminosity distance estimated from the source redshift, $T_j$ the duration of the source emission, $A_{\mathrm{eff}}$ the detector effective area dependent on energy ($E$) and declination ($\delta_j$), $dN/dE$ the energy distribution of signal neutrinos, $E_{\mathrm{min}} = \SI{100}{GeV}$ and $E_{\mathrm{max}} = \SI{10}{PeV}$ the energy range of the injected flux. Outside of this range, the contribution of the flux to the sensitivity can be neglected. The duration $T_j$ is fixed individually for each source according to the duration described in Sec.~\ref{sec:sample}. We note that, for this class of events, neutrino emission models predict long-lasting emissions \cite{Murase:2017pfe, Pitik:2023vcg} compatible with the time scales of the optical evolution. The intrinsic source weight $w^{*}_j$ is proportional to the assumed neutrino luminosity of the source averaged over the time of the emission $T_j$. For the choice of $w^{*}_j$, we adopt two weighting schemes. In the first, we assume a neutrino luminosity proportional to the optical, and define the source weights according to the time-averaged pseudo-bolometric luminosity defined in Sec.~\ref{sec:sample}. In the second scheme, we multiply the average pseudo-bolometric bolometric luminosity by the peak luminosity of the source, as one neutrino emission model \cite{Pitik:2023vcg} favours intrinsecally bright sources. The analysis is implemented using the \texttt{flarestack} \cite{robert_stein_2022_6966491} open source software, that has been purposedly extended with the support for time-dependent searches using Monte Carlo background pseudoexperiments. The sensitivity is defined as the signal intensity at which 90\% of the pseudoexperiments yield a test statistic larger than the background median ($\simeq 0$). For an analysis producing a non-significant result, the sensitivity is conventionally taken as the upper limit to the putative signal. The sensitivity is here reported as the value of the power-law flux normalization at $\SI{1}{GeV}$. We use the first weighting scheme to compare the sensitivity of the "PS" data set to the full catalog and the sensitivities of both datasets, "PS" and "NT", to the Northern sky catalog. These are shown in Fig.~\ref{fig:stacking} for the complete catalog and for the source with the highest weight in the catalog (being the same in the two cases). The shown sensitivities are defined over the energy range contributing for the 90\% to the detector sensitivity, which has been estimated to be between $\SI{3}{TeV}$ and $\SI{5}{PeV}$.
We further consider only the combination of the "PS" sample with the complete catalog and the "NT" sample with the Northern sources selection. We then compare the sensitivities for the two proposed weigthing schemes in Fig.~\ref{fig:stacking-weights}. Here, we divide the total sensitivity by the number of sources in each catalog to obtain an estimate of the sensitivity for an hypothetical source having the weight equal to the catalog average. We find that the "NT" sample provides the best sensitivity.

\section{Ratio of neutrino to optical energy for interaction-powered supernovae}
In the following, we estimate the ratio of the sensitivity flux to the sum of the estimated pseudo-bolometric fluxes of the sources in the catalog. The optical luminosity is a fraction (radiative efficiency) of the kinetic energy powering the shocks \cite{Pitik:2023vcg} that in turn are responsible for particle acceleration. In order to compare neutrino energy and optical energy, we integrate the estimated sensitivity flux over the 90\% sensitive range of the detector obtaining for a spectral index $\gamma=2$:
\begin{equation}
    \eta_{\nu/\mathrm{opt}} = \frac{\phi_0 E_0^2 \log(E_b/E_a)}{\sum_j \frac{L_{\mathrm{opt},j}}{4 \pi d^2_{L,j}}}
\end{equation}
where $\phi_0$ is the sensitivity flux normalization at $E_0 = \SI{1}{GeV}$,  $E_a = \SI{3}{TeV}$ and $E_b = \SI{5}{PeV}$ the integration boundaries, $L_{\mathrm{opt},j}$ the time-averaged pseudo-bolometric luminosity of source $j$ and $d_{L,j}$ its luminosity distance. The best sensitivity estimates of this analysis correspond to a value of $\eta_{\nu/\mathrm{opt}} \simeq 3.5\text{-}3.8$. The model considered in Ref.~\cite{Pitik:2023vcg} evaluates the two most favourable sources, SN2020usa and SN2020in, with flux predictions that correspond to values of $\eta_{\nu/\mathrm{opt}} \sim 0.06\text{-}0.07$, which are significantly below the IceCube sensitivity estimated from this analysis.

\section{Conclusion}\label{sec:conclusion}
We have introduced an analysis aimed to search for an excess of neutrinos from interaction-powered supernovae. We have defined a catalog based on the Bright Transient Survey of the Zwicky Transient Facility, consisting of 74 core-collapse supernovae among type IIn SNe and type II SLSNe. We have estimated the analysis sensitivity for two different IceCube data selections, showing that targeting the sources in the Northern sky leads to the best sensitivity. We propose two weighting schemes, one based on the optical flux of the source and favouring intrinsecally bright sources. We compare the current IceCube sensitivity with predictions from a state-of-the-art theoretical model. We note that the published predictions from such model are limited to the two most promising sources in our catalog, and that the predicted fluxes are significantly below the estimated sensitivity. In light of this, we choose not to adopt a more detailed model-based weighting for individual sources, maintaining the proposed weighting schemes introduced in this work. We plan to update the analysis to include the 2021-22 IceCube data season, that will allow the full coverage of all the detection time windows for the supernovae in our catalogue, before unblinding the results.

\bibliographystyle{ICRC}
\bibliography{references}

%

\clearpage

\input{authorlist_IceCube.tex}

\end{document}

%% file: authorlist_IceCube.tex
\section*{Full Author List: IceCube Collaboration}

\scriptsize
\noindent
R. Abbasi$^{17}$,
M. Ackermann$^{63}$,
J. Adams$^{18}$,
S. K. Agarwalla$^{40,\: 64}$,
J. A. Aguilar$^{12}$,
M. Ahlers$^{22}$,
J.M. Alameddine$^{23}$,
N. M. Amin$^{44}$,
K. Andeen$^{42}$,
G. Anton$^{26}$
% ...
\\
\\
% ...
$^{12}$ Universit{\'e} Libre de Bruxelles, Science Faculty CP230, B-1050 Brussels, Belgium \\
% ...
$^{17}$ Department of Physics, Loyola University Chicago, Chicago, IL 60660, USA \\
$^{18}$ Dept. of Physics and Astronomy, University of Canterbury, Private Bag 4800, Christchurch, New Zealand \\
% ...
$^{22}$ Niels Bohr Institute, University of Copenhagen, DK-2100 Copenhagen, Denmark \\
$^{23}$ Dept. of Physics, TU Dortmund University, D-44221 Dortmund, Germany \\
% ...
$^{26}$ Erlangen Centre for Astroparticle Physics, Friedrich-Alexander-Universit{\"a}t Erlangen-N{\"u}rnberg, D-91058 Erlangen, Germany \\
% ...
$^{40}$ Dept. of Physics and Wisconsin IceCube Particle Astrophysics Center, University of Wisconsin{\textendash}Madison, Madison, WI 53706, USA \\
% ...
$^{42}$ Department of Physics, Marquette University, Milwaukee, WI, 53201, USA \\
% ...
$^{44}$ Bartol Research Institute and Dept. of Physics and Astronomy, University of Delaware, Newark, DE 19716, USA \\
% ...
$^{63}$ Deutsches Elektronen-Synchrotron DESY, Platanenallee 6, 15738 Zeuthen, Germany  \\
$^{64}$ Institute of Physics, Sachivalaya Marg, Sainik School Post, Bhubaneswar 751005, India \\
% ...

\subsection*{Acknowledgements}

\noindent
The authors gratefully acknowledge the support from the following agencies and institutions:
USA {\textendash} U.S. National Science Foundation-Office of Polar Programs,
U.S. National Science Foundation-Physics Division,
U.S. National Science Foundation-EPSCoR,
Wisconsin Alumni Research Foundation,
Center for High Throughput Computing (CHTC) at the University of Wisconsin{\textendash}Madison,
Open Science Grid (OSG),
Advanced Cyberinfrastructure Coordination Ecosystem: Services {\&} Support (ACCESS),
Frontera computing project at the Texas Advanced Computing Center,
U.S. Department of Energy-National Energy Research Scientific Computing Center,
Particle astrophysics research computing center at the University of Maryland,
Institute for Cyber-Enabled Research at Michigan State University,
and Astroparticle physics computational facility at Marquette University;
Belgium {\textendash} Funds for Scientific Research (FRS-FNRS and FWO),
FWO Odysseus and Big Science programmes,
and Belgian Federal Science Policy Office (Belspo);
Germany {\textendash} Bundesministerium f{\"u}r Bildung und Forschung (BMBF),
Deutsche Forschungsgemeinschaft (DFG),
Helmholtz Alliance for Astroparticle Physics (HAP),
Initiative and Networking Fund of the Helmholtz Association,
Deutsches Elektronen Synchrotron (DESY),
and High Performance Computing cluster of the RWTH Aachen;
Sweden {\textendash} Swedish Research Council,
Swedish Polar Research Secretariat,
Swedish National Infrastructure for Computing (SNIC),
and Knut and Alice Wallenberg Foundation;
European Union {\textendash} EGI Advanced Computing for research;
Australia {\textendash} Australian Research Council;
Canada {\textendash} Natural Sciences and Engineering Research Council of Canada,
Calcul Qu{\'e}bec, Compute Ontario, Canada Foundation for Innovation, WestGrid, and Compute Canada;
Denmark {\textendash} Villum Fonden, Carlsberg Foundation, and European Commission;
New Zealand {\textendash} Marsden Fund;
Japan {\textendash} Japan Society for Promotion of Science (JSPS)
and Institute for Global Prominent Research (IGPR) of Chiba University;
Korea {\textendash} National Research Foundation of Korea (NRF);
Switzerland {\textendash} Swiss National Science Foundation (SNSF);
United Kingdom {\textendash} Department of Physics, University of Oxford.